\documentclass[prl,twocolumn]{revtex4}
\usepackage{amssymb}
\usepackage{graphicx}

\begin{document}
\title{Fine structure in the electronic density of states of Al-Ni-Co decagonal quasicrystal from ultrafast time-resolved optical reflectivity } 
\author{T. Mertelj$^{1,2}$, A. O\v slak$^{1}$, J. Dolin\v sek$^{1,2}$, I. R. Fisher$^3$, V.V. Kabanov$^{1}$, D. Mihailovic$^{1,2}$}
\author{}
\date{\today}

\begin{abstract}

We measured the temperature and fluence dependence of the time-resolved photoinduced optical reflectivity in a decagonal Al$_{71.9}$Ni$_{11.1}$Co$_{17.0}$ quasicrystal. We find no evidence for the relaxation of a hot thermalized electron gas as observed in metals. Instead, a quick diffusion of the hot nonthermal carries $\sim$ 40 nm into the bulk is detected enhanced by the presence of a broad $\sim$1 eV pseudogap. From the relaxation dynamics we find an evidence for the presence of a fine structure in the electronic density of states around $\sim$13 meV from the Fermi energy. The structure is related to a weak bottleneck for the carrier relaxation observed at low temperatures.

\end{abstract}

\affiliation{$^{1}$Jozef Stefan Institute, Jamova 39, 1000
Ljubljana, Slovenia}

\affiliation{$^{2}$Faculty of Mathematics and Physics, University of
Ljubljana, Jadranska 21, 1000 Ljubljana, Slovenia}

\affiliation{$^{3}$ Geballe Laboratory for Advanced Materials, Dept. of Applied Physics, Stanford University, California 94305, USA}

\maketitle

The ultrafast relaxation of hot photoexcited electrons in systems with suppressed density of electronic states (DOS) near the the Fermi energy ($E_\mathrm F$) such as semimetallic graphite\cite{SeibertCho1990}, the cuprates\cite{HanVardeny1990} and charge-density wave systems\cite{DemsarBiljakovic1999} has received a lot of attention in the past two decades. The presence of a gap, or a pseudogap near $E_\mathrm F$ leads to a bottleneck in the hot-carriers relaxation and the emission of hot optical phonons.\cite{KabanovDemsar99,KampfrathPerfetti2005}

A suppression of the DOS near $E_\mathrm F$ is characteristic also for quasicrystals (QC).\cite{StadnikPurdie97} The pseudogap is narrow in icosahedral QC and of the order of $\sim$1eV wide in decagonal QC (d-QC). However, there is some evidence that a finer structure might exist in the DOS near $E_\mathrm F$ also in d-QC.\cite{OkadaEkino2007} In addition d-QC are characterized by strong anisotropies of properties along periodic axis and quasiperiodic (QP) plane.\cite{BianchiBommeli98} It is therefore an interesting fundamental question what are the relaxation pathways of hot photoexcited electrons in d-QC which show metallic conductivity along the periodic direction and nonmetallic one in the quasiperiodic plane\cite{MartinHebard1991}. In this Letter we present results of optical pump-probe spectroscopy in a decagonal Al$_{71.9}$Ni$_{11.1}$Co$_{17.0}$ QC. From the relaxation dynamics we find evidence for the presence of a structure in the DOS on a scale of $\sim$13 meV around the Fermi energy and no evidence for the relaxation of a hot \emph{thermalized} electron gas\cite{Allen87}

The preparation and characterization of a single grain decagonal Al$_{71.9}$Ni$_{11.1}$Co$_{17.0}$ quasicrystal was described elsewhere.\cite{FisherKramer1999}. Optical measurements were performed on a side facet, which is parallel to the 10-fold axis of the prism shaped sample mounted in an optical cryostat.  A standard pump-probe setup was used with linearly polarized pump beam with the photon energy 1.55 eV, the pulse length 50 fs and repetition frequency 250 kHz. The pump beam was focused to a 250-$\mu$m diameter spot on the facet in a nearly perpendicular geometry. To detect the photoinduced reflectivity $\Delta R/R$ a weaker probe beam with the photon energy, $\hbar \omega_{probe}$,  either 1.55 or 3.1 eV and the diameter 220 $\mu$m was focused to the same spot and detected upon reflection by a PIN photodiode.

\begin{figure}[h]
  \begin{center}
  \includegraphics[angle=-90,width=0.47\textwidth]{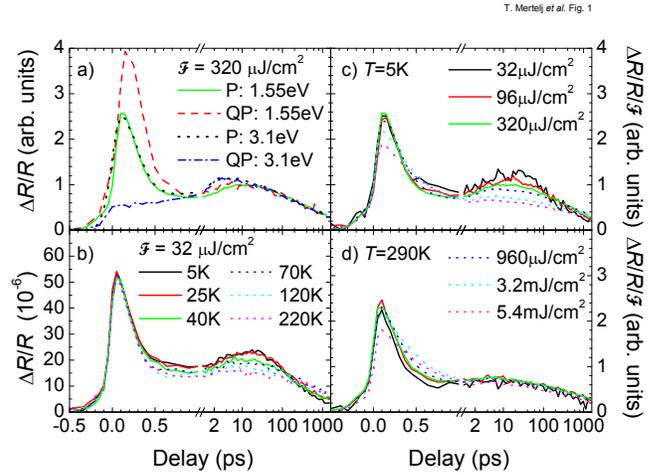}
  \end{center}
  \caption{The probe polarization and photon energy dependence of photoinduced-reflectivity transients at 5K (a). The scans are normalized to enable easier comparison. Temperature dependence of photoinduced-reflectivity transients (b) and fluence dependence of photoinduced-reflectivity transients at two different temperatures (c) and (d) at 1.55-eV probe photon energy and polarization along the periodic direction  .}
\end{figure}

In Fig. 1a we show the dependence of the $\Delta R/R$ transients on the probe polarization at two different photon energies at the temperature 5K. We can identify three distinct timescales on which the transients show different behaviour. On the sub-ps timescale one can clearly identify a sub-ps component with fast $\sim$200-fs rise and decay times. The component is present in all traces except for the polarization along QP direction at 3.1eV-PPE where another component is revealed, which slowly rises on a few-ps timescale after initial $\sim$100-fs rise. At an intermediate timescale between $\sim$1 and $\sim$20 ps we observe further rise of the $\Delta R/R$ which only starts to relax towards equilibrium after $\sim$2 ps with 3.1-eV PPE and after $\sim$10 ps with 1.55- PPE. The dispersion of the $\Delta R/R$ rise indicates a two component relaxation on this timescale as well. On a long timescale beyond $\sim$20 ps all curves, when properly scaled, fall onto each other indicating a common origin of the relaxation. 

We should also note that we observed no anisotropy with respect to the 1.55-eV pump-beam polarization orientation on all timescales.

To study further the temperature and pump fluence ($\cal{F}$) dependence we chose 1.55-eV PPE and the polarization along the periodic direction which has the largest signal to noise ratio and contains all relaxation components. In Fig. 1b we show the temperature dependence of the transients at the lowest $\cal{F}$. The sub-ps component is clearly temperature independent, while at the intermediate timescale a shift of the $\Delta R/R$ peak towards shorter delay and decrease of the peak amplitude are observed with an increasing temperature.

The $\cal{F}$ dependence at 5K (Fig. 1c) is somewhat similar to the temperature dependence, except that at the highest two $\cal{F}$ the sub-ps component shows saturation and an increase of the relaxation time. At 290K (Fig. 1d) the increase of the $\cal{F}$ only influences the sub-ps component, while the relaxation beyond $\sim$1 ps is $\cal{F}$-independent.

\begin{figure}[h]
  \begin{center}
  \includegraphics[angle=-0,width=0.25\textwidth]{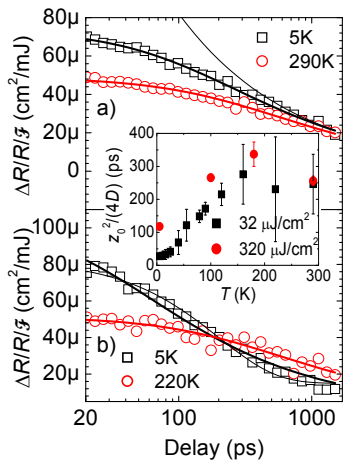}
  \end{center}
  \caption{Fits of equation (1) to $\Delta R/R$ at $\cal{F}$ = 320 $\mu$J/cm$^2$ (a) and at  $\cal{F}$ = 32 $\mu$J/cm$^2$ (b). For comparison a function proportional to $1/\sqrt{t}$ in (a) and a single exponential fit in (b) are shown by  thin solid lines.  The inset shows the temperature dependence of the fit parameter.}
\end{figure}

We start the analysis by considering the long timescale first. The universal delay dependence at different polarizations and PPE indicates that on this timescale all the subsystems are in a local thermal equilibrium and can be described by a single local temperature. The time dependence of the local temperature is governed by the 1D heat diffusion equation. Since the optical penetration depth at the probe photon energies is estimated\footnote{We determined optical constants from angular dependence of the reflectivity.} to be $\sim$ 25 nm, we assume that $\Delta R/R$ is in the lowest order proportional to the surface temperature rise $\Delta T_s(t)$.

If, after a local temperature is established, the energy is distributed within the surface layer of depth $z_0$, the time dependence of the surface temperature is given by: 
\begin{equation}
\Delta T_s(t)=\frac{\Delta T_0}{\sqrt{1+ \frac{4 D t}{z_0^2}}.},
\end{equation}
where $D$ represents the heat diffusivity along QP direction in our experimental geometry.\footnote{For simplicity we assume that the initial temperature is given by $\Delta T_0(z)=\Delta T_0\exp (-z^2/z_0^2)$ and that $D$ is $T$ independent.} Equation (1) fits the measured $\Delta R/R$ on the long timescale rather well(see Fig. 2) except for low $\cal{F}$ at low $T$, where the temperature dependence of $D$ and nonlinear dependence of $\Delta R/R$ on $\Delta T_s$  become important. From the result of the fits we estimate the energy deposition depth $z_0$ and the initial temperature rise $\Delta T_0$  using published values of the heat capacity\cite{BarrowCook2003,InabaLortz2002} and thermal conductivity along the QP direction\cite{BianchiBommeli98,BarrowCook2003}. The estimated $z_0$ shown in Fig. 3a first increases and then drops slowly with increasing $T$ to $\sim$ 40 nm at 290K. Although we obtain at 5K virtually the same $z_0$ from measurements at two different $\cal{F}$ the values below $\sim$100K should be taken with caution due to the limited validity of equation (1). 

Surprisingly, at all temperatures $z_0$ is significantly larger than the estimated optical penetration depth of $\sim$ 25 nm indicating a quick diffusion of hot carriers into the sample during the first few picoseconds of relaxation. This is rather unexpected due to the strong Drude damping in the QP plane\cite{BianchiBommeli98}. One can roughly estimate the hot electron diffusion constant by neglecting the energy transfer to the lattice, $D_\mathrm{el}=\lambda_\mathrm{el}/c_\mathrm{el}\rho$, where $\lambda_\mathrm{el}$ is the electronic heat conductivity, $c_\mathrm{el}$ the electronic heat capacity and $\rho$ the mass density. We calculate $\lambda_\mathrm{el}$ from resistivity\cite{FisherKramer1999}, which is almost $T$-independent, using Wiedemann-Franz's law. By taking $c_\mathrm{el}$ from\cite{BianchiBommeli98} we obtain $D_\mathrm{el}\simeq 1$ cm$^2$/s, which is virtually temperature independent. In 1 ps this gives 20 nm diffusion length which is of correct magnitude to explain observations.

\begin{figure}[h]
  \begin{center}
  \includegraphics[angle=-0,width=0.25\textwidth]{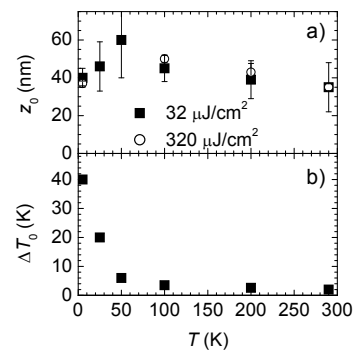}
  \end{center}
  \caption{The temperature dependence of the initial energy-distribution depth (a) and the initial surface-temperature rise at 32-$\mu$J/cm$^2$ excitation fluence(b).}
\end{figure}

We proceed with the analysis of the intermediate-timescale. Temperature dependencies of the peak amplitude and the intermediate-timescale rise time shown in Fig. 4 clearly suggest that there exists a well defined energy scale of the order of $\sim$100K associated with the processes occurring on this timescale. We assume that $\Delta R$ is a consequence of the excited state absorption\cite{DvorsekKabanov2002},
\begin{equation}
\Delta R \propto \int d\epsilon N(\epsilon)|M(\epsilon,\omega)|^2[f(\epsilon)-f_T(\epsilon)],
\end{equation}
where $\hbar\omega$ is the energy of the photons, $\epsilon$ the energy measured from the Fermi energy, $N(\epsilon)$ is the effective density of the initial states close to the Fermi energy\footnote{We assume that on the timescale of a few ps the final states, which are $\hbar\omega$ away from the Fermi energy are completely unoccupied and that the optical resonance is broad with respect to $k_ \mathrm B T$.}, $M(\epsilon,\omega)$ the effective dipole transition matrix element, $f(\epsilon)$ the non-equilibrium electron distribution function and $f_T(\epsilon)$ the equilibrium Fermi function. If $N(\epsilon)$ has in addition to a smooth background a narrow peak at the energy $E_p$ there will exist a contribution $\Delta R_\mathrm p$ to $\Delta R$ proportional to $f(E_\mathrm p)-f_T(E_\mathrm p)$ in addition to a weakly temperature dependent contribution  $\Delta R_\mathrm b$ from the smooth background. 

In the simplest case one can assume that the nonequilibrium  distribution is thermal, $f(E_\mathrm p)=f_{T'}(E_\mathrm p)$, at an elevated temperature which is the same as the lattice temperature $T'=T+\Delta T_0$, where $\Delta T_0$ is estimated form the long timescale behaviour. In Fig. 4a we show the difference $\Delta f_{T}(E_\mathrm p)=f_{T'}(E_\mathrm p)-f_T(E_\mathrm p)$ for three different $E_\mathrm p$ using values of $\Delta T_0$ from Fig. 3b. It is clearly seen that $\Delta f_{T}(E_\mathrm p)$ at $E_\mathrm p/k_ \mathrm B=$ 150K reproduces the measured temperature dependence rather well if one assumes that $\Delta R_\mathrm b$ is temperature independent.

\begin{figure}[h]
  \begin{center}
  \includegraphics[angle=-0,width=0.3\textwidth]{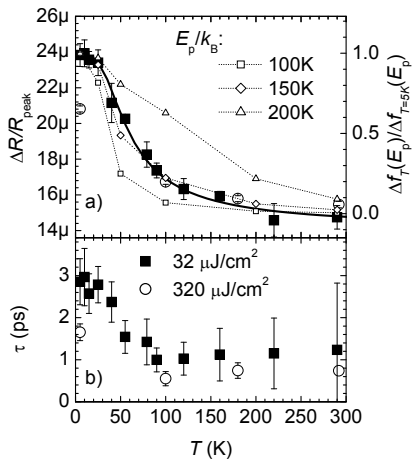}
  \end{center}
  \caption{Temperature dependencies of the intermediate-timescale peak magnitude of $\Delta R/R$ (a) and  the intermediate-timescale rise time (b) at the two different fluences. In (a) the small open symbols connected with dotted lines represent $\Delta f_{T}(E_\mathrm p)$ discussed in text.}
\end{figure}

The alternative scenario is the existence of a bottleneck due to the suppressed density od states near the Fermi energy similar to the pseudogap in the cuprate superconductors. In this case for small excitation densities and strong bottleneck the density of the photoexcited quasiparticles is given by\cite{KabanovDemsar99},
\begin{equation}
n_\mathrm{pe} \propto  1/[1+B \exp(-\Delta_\mathrm g/k_\mathrm B T)],
\end{equation} 
where $\Delta_\mathrm g$ represents the pseudogap width and $B$ the ratio between number of phonon degrees of freedom and number of electronic states in the phonon-energy range. A fit of equation (3) to the data is shown in Fig. 4a. As in the previous case we assumed that $\Delta R$ is a sum of a temperature independent part $\Delta R_\mathrm b$ and a part proportional to equation (3). The resulting pseudogap width is virtually the same as in the previous case, $\Delta_\mathrm g/k_\mathrm B=$140K$\pm$40K. The large error bar comes from indeterminacy of $\Delta R_\mathrm b$.

\begin{figure}[h]
  \begin{center}
  \includegraphics[angle=-0,width=0.3\textwidth]{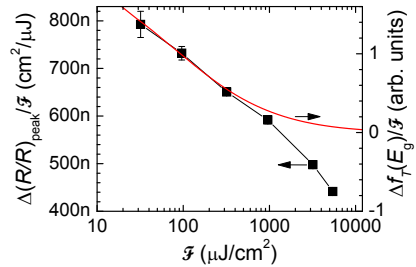}
  \end{center}
  \caption{Fluence dependence of the intermediate-timescale peak magnitude of $\Delta R/R$ at 5K. For comparison $\Delta f_{T}(E_\mathrm p)$ is shown with $E_\mathrm p/k_ \mathrm B=$ 150K as the solid line assuming 45-nm initial energy deposition depth and published values of the heat capacity data\cite{BianchiBommeli98,BarrowCook2003,InabaLortz2002} and reflectivity\cite{BianchiBommeli98} in similar decagonal quasicrystals.}
\end{figure}

Both models are based on an increased density of states at $\sim$13 meV away from the Fermi energy and are just two limiting cases representing either no or a strong bottleneck. In the strong bottleneck case $\Delta R_{peak}/R$ is proportional to $\cal{F}$ at small $\cal{F}$ while at large $\cal{F}$ a crossover to $\Delta R_{peak}/R \propto \sqrt{\cal{F}}$ is expected.\cite{KabanovDemsar99} We do not observe any of the dependencies in our case. On the other hand, the $\cal{F}$ dependence of $\Delta R_{peak}/R$ for small $\cal{F}$ in the no-bottleneck model is compatible with the experimental one shown in Fig. 5. Unfortunately, the no-bottleneck model fails to fit the data beyond $\cal{F}\approx$ 1 mJ/cm$^2$ as well. The reason for this might be in $\cal{F}$-dependence of $\Delta R_\mathrm b$, which also shows some saturation beyond $\cal{F}\approx$ 3 mJ/cm$^2$ judging from the measurements at 290K.(see Fig. 1d).

While the $\cal{F}$-dependence clearly excludes the strong bottleneck scenario it does not exclude a weak bottleneck, the presence of which is suggested by the increase of the intermediate-scale risetime below 100K (see Fig 4b). In order to elaborate this we need to analyze the response on the shortest timescale first. 

The sub-ps peak amplitude has linear $\cal{F}$-dependence and the decay time has no $\cal{F}$-dependence up to $\cal{F} \approx $ 3 mJ/cm$^2$. In addition, both show no $T$-dependence so the sub-ps dynamics cannot be attributed to the thermal relaxation\cite{Allen87} where $f(\epsilon)$ is assumed to be the Fermi distribution with the nonequilibrium electronic temperature $T_\mathrm e$. $f(\epsilon)$ is therefore nonthermal on the sub-ps timescale and the sub-ps peak is due to the relaxation across the broad pseudogap\cite{KrajciHafner2000,StadnikPurdie97}  and the hot carriers diffusion into the sample. The broad pseudogap suppresses the probability for the creation of the low energy electron-hole (e-h) pairs increasing the electron thermalization time beyond $\sim$500 fs enhancing the hot carrier diffusion into the sample.

There is also no evidence for the thermal relaxation of the hot electron gas on a timescale up to several ps. The intermediate-timescale risetime increases with decreasing temperature which, is inconsistent with the thermal relaxation\cite{Allen87}, where the relaxation time is predicted to be proportional to the lattice temperature $T_\mathrm L$ for $T_\mathrm e \sim T_\mathrm L$.
The intermediate-timescale dynamics might therefore at least in part be attributed to the dynamic lattice expansion due to an increasing lattice temperature\cite{RichardsonSpicer2002}. The expansion, however, also cannot explain  the increase of the intermediate-timescale risetime at low temperatures,  implying another relatively slow relaxation channel. We tentatively assign this channel to the relaxation of hot nonthermal optical phonons. We believe that due to the increased thermalization time a significant number of hot optical phonons is generated in addition to the low energy e-h pairs during the thermalization. These nonthermal hot optical phonons can decay by the anharmonic decay or by excitation of the e-h pairs. At low temperatures the anharmonic decay channel is suppressed resulting in a bottleneck and the increase of the intermediate-timescale risetime below 100K.

The relaxation upon photoexcitation in Al$_{71.9}$Ni$_{11.1}$Co$_{17.0}$ quasicrystal therefore proceeds roughly in two steps. In the first step the presence of the broad pseudogap slows down the hot carrier thermalization. During the relatively slow thermalization hot carriers diffuse up to $\sim$40 nm into the bulk and a part of the absorbed energy goes to the nonthermal hot phonons.  During the second step, which is at 5K characterized by the relaxation time of $\sim$3 ps, the hot nonthermal phonons decay. At low temperatures the nonthermal-phonon decay is suppressed due to the decrease of the anharmonic relaxation rate and the weak bottleneck in the relaxation through the electronic channel. The bottleneck is a consequence of a fine structure in the electronic DOS at $\sim$13 meV from the Fermi energy.

\begin{acknowledgments}

\end{acknowledgments}

%\begin{thebibliography}
\bibliography{anc}
%\end{thebibliography}
\end{document}